\RequirePackage[displaymath]{lineno}
\documentclass[twocolumn,aps,prc,showpacs,superscriptaddress,floatfix,nofootinbib]{revtex4}
\usepackage{newtxtext,newtxmath,booktabs,siunitx}
\usepackage{dcolumn}% Align table columns on decimal point
\usepackage{bm}% bold math
\usepackage{float}
\usepackage[]{graphicx}  % remove 'demo' option for your real document
\usepackage{booktabs}
\usepackage{tabularx}
\usepackage{amsmath}
\usepackage{xcolor}
\usepackage{multirow}
\usepackage[colorlinks,citecolor=blue,urlcolor=blue,linkcolor=blue]{hyperref}

\newcommand {\snn}      {\sqrt{s_{_{\rm NN}}}}

\newcommand {\Nch}      {N_{\rm ch}}

\newcommand {\Zr}       {$^{96}$Zr}
\newcommand {\Ru}       {$^{96}$Ru}
\newcommand {\RuRu}     {$^{96}_{44}$Ru+$^{96}_{44}$Ru}
\newcommand {\ZrZr}     {$^{96}_{40}$Zr+$^{96}_{40}$Zr}

\newcommand {\eccn}     {\epsilon_{n}}
\newcommand {\vn}       {v_{n}}

\newcommand {\mean}[1]  {\langle #1\rangle}

\newcommand {\RNch}  {R_{\mean{\Nch}}}
\newcommand {\Ren}  {R_{\epsilon_{2}}}
\newcommand {\Reo}  {R_{\epsilon_{3}}}
\newcommand {\Reno}  {R_{\epsilon_{n}}}

\begin{document}
\title{Impact of initial fluctuations and nuclear deformations in isobar collisions}

\author{Jian-fei Wang}
\affiliation{Department of Modern Physics, University of Science and Technology of China, Anhui 230026, China}
\affiliation{School of Science, Huzhou University, Huzhou, Zhejiang 313000, China}
\author{Hao-jie Xu\footnote{%
        Corresponding author: haojiexu@zjhu.edu.cn}}
\affiliation{School of Science, Huzhou University, Huzhou, Zhejiang 313000, China}
\affiliation{Strong-Coupling Physics International Research Laboratory (SPiRL), Huzhou University, Huzhou, Zhejiang 313000, China.}
\author{Fuqiang Wang}
\affiliation{School of Science, Huzhou University, Huzhou, Zhejiang 313000, China}
\affiliation{Strong-Coupling Physics International Research Laboratory (SPiRL), Huzhou University, Huzhou, Zhejiang 313000, China.}
\affiliation{Department of Physics and Astronomy, Purdue University, West Lafayette, Indiana 47907, USA}

\begin{abstract}
Relativistic isobar (\RuRu\ and \ZrZr) collisions have revealed intricate  differences in their nuclear size and shape, inspiring unconventional studies of nuclear structure using relativistic heavy ion collisions. 
In this study, we investigate the relative differences in the mean multiplicity ($\RNch$) and the second- ($\Ren$) and third-order eccentricity ($\Reo$) between isobar collisions using initial state Glauber models. 
	It is found that initial fluctuations and nuclear deformations have negligible effects on $\RNch$ in most central collisions, while both are important for the $\Ren$ and $\Reo$, the degree of which is sensitive to the underlying nucleonic or sub-nucleonic degree of freedom. These features, compared to real data, may probe the particle production mechanism and the physics underlying nuclear structure.
\end{abstract}
\maketitle

\section{Introduction}
The recent isobar data~\cite{STAR:2021mii,STAR:2019bjg} stimulated a wide interest from the relativistic heavy ion community in the physics of nuclear structures of isobars (nuclei with equal number of nucleons but different numbers of protons and neutrons)~\cite{Xu:2017zcn,Li:2018oec,Li:2019kkh,Zhang:2021kxj,Nijs:2021kvn,Jia:2022qgl,Zhang:2022fou,Nie:2022gbg,Xi:2023isk,Cheng:2023ucp}. 
The isobar collisions of \RuRu\ and \ZrZr\ were initially proposed to search for the chiral magnetic effect (CME)~\cite{Kharzeev:2007jp,Voloshin:2010ut,Deng:2016knn,Wang:2018ygc}.
The measurements in isobar collisions at $\snn=200$ GeV by the STAR Collaborations showed sizable differences in the multiplicity distribution, elliptic flow, and triangular flow~\cite{STAR:2021mii}, indicating  differences in the density profiles of the colliding nuclei~\cite{Xu:2017zcn,Zhang:2021kxj}. 
Nuclear density is usually obtained from theory calculations, such as energy density functional theory (DFT)~\cite{Chabanat:1997qh,Chamel:2009yx,Wang:2016rqh,Zhang:2015vaa}, incorporating information from experimental measurements. In relativistic heavy ion collision simulations, it is common to use the parameterized nuclear density in the Woods-Saxon formula, 
\begin{equation}
\rho(r) = \frac{\rho_{0}}{\exp{\left[\left(r-R(1 + \beta_{2}Y{2}^{0} + \beta_{3}Y_{3}^{0}+...\right)/a\right]}+1}, 
\end{equation}
with the radius parameter $R$, the diffuseness parameter $a$, and the quadrupole (octupole) deformation parameter $\beta_{2}$ ($\beta_{3}$).
A large $a$ value in \Zr\ is suggested by the STAR data on multiplicity distribution and elliptic flow ratios in non-central collisions~\cite{STAR:2021mii}, consistent with a halo-type neutron skin thickness as predicted by the DFT calculations~\cite{Xu:2021vpn}.
The flow  ratios observed in the most central collisions suggest that the \Ru\ has a larger $\beta_{2}$, while \Zr\ has a larger $\beta_{3}$~\cite{Zhang:2021kxj}.

Much progress has been made in quantitatively assessing the differences in size and shape of isobars by using relativistic isobar collisions~\cite{Li:2019kkh,Xu:2021qjw,Xu:2021uar,Xu:2022ikx,Liu:2022kvz,Nijs:2021kvn,Jia:2022qgl,Zhang:2022fou,Cheng:2023ucp,Luzum:2023gwy}. %These unconventional ways of nailing down the differences in the isobar structure parameters can, in turn, 
These differences can place stringent constraints on key parameters in theoretical modeling of nuclear interactions, which is the backbone of all theoretical calculations of nuclear structure including DFT. One of such parameters is the slope parameter $L(\rho)$ of the nuclear symmetry energy as a function of nuclear density, which cannot yet be well determined by low-energy nuclear experiments~\cite{Chen:2005ti,Tsang:2012se,Cai:2014kya,Li:2016gjs,Yu:2020epq,Li:2021thg,Wang:2022cda,Ma:2022lox,An:2023ahu}. In addition, nuclear deformations are largely uncertain from nuclear structure calculations~\cite{Moller:2015fba,DRHBcMassTable:2022uhi,Deng:2016knn}. 
On the other hand, the STAR experiment has collected approximately 2 billion minimum-bias events for each species of \RuRu\ and \ZrZr\ collisions at $\snn=200$ GeV, with good control of systematic uncertainties~\cite{STAR:2021mii}. 
These provide an unique opportunity to study the density slope parameter of the symmetry energy and the deformation parameters of the isobars.
Previous studies suggest that ratio observables in isobar collisions are less sensitive to  final-state interactions~\cite{Li:2019kkh,Xu:2021vpn,Zhang:2022fou}. In this work, we investigate the effect of  initial-state fluctuations on those ratio observables in relativistic isobar collisions.

We use the Optical Glauber model and Monte Carlo Glauber model~\cite{Kolb:2000sd,Miller:2007ri,Loizides:2017ack} to investigate the effect of nuclear deformations and initial fluctuations on the relative differences $\RNch$, $\Ren$, and $\Reo$ in relativistic isobar collisions, where
\begin{equation}
R_{X} \equiv 2 \frac{X_{\rm RuRu} - X_{\rm ZrZr}}{X_{\rm RuRu} + X_{\rm ZrZr}}.
\end{equation}
The Glauber model is widely used to determine the centrality in experiments, including the blind analysis of the STAR isobar data~\cite{STAR:2021mii}. 
Due to symmetry, the elliptic flow $v_{2}$ with impact parameter $b=0$ and the triangular flow $v_{3}$ can not be generated in spherical nucleus-nucleus collisions with smooth initial conditions in Optical Glauber simulation. The collision configuration of deformed nuclei will contribute to spatial anisotropy, while event-by-event fluctuation contributions are also important.
We further investigate the effect of initial fluctuations by comparing the results under the scenarios of nucleon participants and quark participants. 

The rest of the paper is organized as follows. Section~\ref{sec:model} gives a brief description of the models and the nuclear density profiles used in this work. Section~\ref{sec:results} discusses the effects of nuclear deformation and initial fluctuations on $\RNch$, $\Ren$, and $\Reo$. A summary is given in Sec.~\ref{sec:summary}.

\section{Glauber model and isobar nuclear densities}\label{sec:model}
Particle production in relativistic heavy ion collisions is usually parameterized by the two-component model, 
\begin{equation}\label{eq:2compt}
\Nch = n_{\rm pp}[(1-x)N_{\rm part}/2 + xN_{\rm coll}], 
\end{equation}
with the number of participants $N_{\rm part}$ and number of binary collisions $N_{\rm coll}$ obtained from the Glauber model~\cite{Miller:2007ri}.
Here the $x$ is the relative contribution of hard processes and $n_{\rm pp}=\Nch$ in nucleon-nucleon collisions. 
For the case of $x=0$, the particle production is independent of hard processes and it can alternatively be described by a variant of the Glauber model, the so-called Trento model~\cite{Moreland:2014oya,Bernhard:2016tnd}, 
\begin{equation}
\Nch \propto \int s(r_{\perp})d^2r_{\perp}\propto \int \left( \frac{T_{A}^{p}(r_{\perp})+T_{B}^{p}(r_{\perp})}{2} \right)^{1/p}d^2r_{\perp}
\end{equation}
with $p=1$ (we note that previous studies favor $p\sim0$~\cite{Bernhard:2016tnd}). 
Here $s$ is the entropy density, $r_{\perp}$ is the transverse radius, and $T_{A(B)}$ is the reduced thickness function for the colliding nuclei $A(B)$ with $\int (T_{A} + T_{B})d^2r_{\perp} = N_{\rm part}$. In this work, we focus on the centrality dependence of the ratio observables in isobar collisions, so the parameter $n_{\rm pp}$ and the normalization factor hidden in the above equation are not relevant to our study.

The elliptic flow ($v_{2}$), triangular flow ($v_{3}$) are the  Fourier coefficients of the azimuthal angle ($\phi$) distribution
$\vn = \langle \cos\left[n\left(\phi-\Psi_{n}\right)\right]\rangle$,
 where $\langle...\rangle$ denotes the average over an event and  $\Psi_{n}$ is the plane angle in momentum phase. The large anisotropic flow  can be successfully described by relativistic hydrodynamics with a large spatial eccentricity~\cite{Poskanzer:1998yz},
\begin{equation}
        \eccn e^{in\Phi_{n}}  =\frac{\int r_{\perp}^{n}e^{in\varphi}w(r_{\perp})d^2r_{\perp}}{\int r_{\perp}^{n} w(r_{\perp})d^2r_{\perp}}.
\end{equation}
Here $w(r_{\perp})$ is the weight factor, often taken to be the entropy density $s$ or the energy density $e$. $\Phi_{n}$ is the $n$th order plane angle in the configuration space. 
The anisotropic flow is approximately
proportional to the initial eccentricity for small amplitude~\cite{Noronha-Hostler:2015dbi}. 
In this work, we use the participant density to calculate $\epsilon_{2}$ and $\epsilon_{3}$, however, our main conclusions do not change if alternative $w$ is used. 

For spherical nucleus-nucleus collisions, the multiplicity and eccentricities can be calculated analytically in the Optical Glauber model for a given impact parameter $b$. For deformed nucleus-nucleus collisions, we calculate the ratio observables at a given shape orientation with a specific Euler rotation and then take the average over all orientations 
\begin{equation}
    \mean{X(b)} = \frac{1}{(4\pi)^2} \int X(\Omega_A,\Omega_B; b) d\Omega_Ad\Omega_B.
\end{equation}
Here $X$ is the observable of interest and  $\Omega_{A(B)}$ is the orientation solid angle for colliding nuclei A(B). 

The event-by-event fluctuations are not included in the Optical Glauber model~\cite{Miller:2007ri}, and such an effect is crucial for the anisotropic flow observables, especially the odd-order ones. We, therefore, use the Monte Carlo Glauber model to include the fluctuation effect, where the finite number of nucleons ($96$ for the Ru and Zr isobars) are sampled in each colliding nucleus. The event-by-event fluctuations depend on the number of nucleons in the colliding nuclei, so these effects are expected to be larger in the isobar collisions than in Au+Au collisions. After sampling the nucleons with the appropriate nuclear density, the multiplicity is calculated based on Eq.~(\ref{eq:2compt}), and the eccentricity is calculated from the transverse position of those finite participating nucleons. 
Here the participants and binary collisions are determined 
from nucleon-nucleon interactions which happens when
the distance between the two nucleons, each from the incoming and outgoing nuclei, is less than $\sqrt{\sigma_{NN}/\pi}$~\cite{Miller:2007ri},
with the inelastic cross section set to $\sigma_{\rm NN}=42$ mb for $\snn=200$ GeV.
Poisson-type multiplicity fluctuations are applied in the Monte Carlo simulations.

\begin{table}
	\caption{Woods-Saxon parameterizations of the \Ru\ and \Zr\ nuclear density distributions. The quoted values for $R_0$ and $a$ are in fm and that for $\rho_0$ is in 1/fm$^3$. \label{tab:WSDFT}}
	\centering{}%
    \begin{tabular}%{ccccccc}
	    {p{1.0cm}p{0.9cm}p{0.9cm}p{0.9cm}|p{1.0cm}p{0.9cm}p{0.9cm}p{0.9cm}}
    \hline
 %   {p{0.9cm}p{0.9cm}p{0.9cm}p{0.9cm}|p{0.9cm}p{0.9cm}p{0.9cm}p{0.9cm}}
      \multicolumn{4}{c|}{\Ru} & \multicolumn{4}{c}{\Zr}  \\
	   $\rho_{0}$ & $R$    & $a$   & $\beta_2$ & $\rho_{0}$ & $R$ & $a$ & $\beta_3$  \\ 
    \hline
	    0.159 & 5.093 & 0.487 & 0.00 & 0.163 & 5.022 & 0.538 & 0.00  \\
	    0.159 & 5.093 & 0.471 & 0.16 & 0.163 &5.021  & 0.517  & 0.20  \\
	\hline
	\end{tabular}
\end{table}

\begin{table*}

	\caption{The fixed impact parameter used to mimic the 16 equal-size centrality bins from $0-80\%$ centrality range. \label{tab:Opt}}
	\centering{}%
    \begin{tabular}%{ccccccccccccccccc}
	{p{1.8cm}p{0.8cm}p{0.8cm}p{0.8cm}p{0.8cm}p{0.8cm}p{0.8cm}p{0.8cm}p{0.8cm}p{0.8cm}p{0.8cm}p{0.8cm}p{0.8cm}p{0.8cm}p{0.8cm}p{0.8cm}p{0.8cm}}
    \hline
	  Centrality ($\%$) & 2.5 & 7.5  & 12.5 & 17.5 & 22.5 & 27.5 & 32.5 & 37.5 & 42.5 & 47.5  
   & 52.5 & 57.5 & 62.5 & 67.5 & 72.5 & 77.5 \\ 
    \hline
	   $b_{\rm Ru}$ (fm)  & 1.865 & 3.230 & 4.170 & 4.933 & 5.594 & 6.184 & 6.723 & 7.222 & 7.688 &
      8.128 & 8.545 & 8.943 & 9.323 & 9.689 & 10.042 & 10.384\\
 	   $b_{\rm Zr}$ (fm)  & 1.903 & 3.296 & 4.255 & 5.035 & 5.709 & 6.312 & 6.862 & 7.370 & 7.846 &
      8.295 & 8.721 & 9.127 & 9.515 & 9.888 & 10.249 & 10.600\\
	\hline
	\end{tabular}
\end{table*}

The density profiles of \Ru\ and \Zr\ nuclei are required in the Glauber simulations. 
In this study, the parameters for the Woods-Saxon density distributions 
of \Ru\ and \Zr\ are taken from Ref.~\cite{Xu:2021qjw,Zhao:2022uhl} (see Tab.~\ref{tab:WSDFT}), which are extracted from the isobar densities obtained by DFT calculations with the symmetry energy $E_{\rm sym}(\rho_{c})=26.65$ MeV and its density slope parameter  $L(\rho_{c})=47.3$ MeV at subsaturation cross density $\rho_{c}\sim0.11$ fm$^{-3}$~\cite{Zhang:2015vaa}. Nuclear deformation is not included in the DFT calculations due to its strong model dependence. Following our previous work~\cite{Xu:2021qjw,Zhao:2022uhl}, to introduce the nuclear deformation, the WS parameters $R$ and $a$ for the given $\beta_{2}$ (or $\beta_{3}$) are calculated to match the volume and root-mean-square (RMS) of the corresponding nucleus calculated by DFT, keeping the normalization factor $\rho_{0}$ fixed. 

Centrality is determined by the multiplicity distributions in the Monte Carlo Glauber model, similar to what was  done in heavy ion experiments~\cite{STAR:2021mii}. For the optical Glauber model and optical Trento model in spherical nucleus-nucleus collisions, the centrality is related to the differential cross-section in impact parameter~\cite{Miller:2007ri}. In this study, we use the 16 equal-size centrality bins from $0-80\%$ centrality range. The results obtained with the fixed impact parameter corresponding to the centre of each centrality bin are used to mimic the observables in such a centrality bin. The impact parameters for \Ru\ and \Zr\ are listed in Tab.~\ref{tab:Opt}. The impact parameters are larger in \ZrZr\ collisions than in \RuRu\ collisions for a given centrality bin because the total cross section for \ZrZr\ collisions ($\sigma_{X}^{\rm ZrZr}=4.551$ barn) is larger than that for \RuRu\ collisions ($\sigma_{X}^{\rm RuRu}=4.369$ barn) based on the nuclear density parameters listed in Tab.~\ref{tab:WSDFT}.  For simplicity, the impact parameter list is also used to determine the centrality for deformed nucleus-nucleus collisions in Optical Glauber simulations. We note that determining centrality in the deformed case is in principle more complicated, but this does not  affect our conclusion (see discussion in the next section).

In order to achieve high precision of calculations in our study, the GPU parallel computing technology~\cite{Wu:2019tsf} is used in our Glauber model simulations.

\section{Results and discussions}\label{sec:results}

\begin{figure*}[hbt!]
      \includegraphics[scale=0.4]{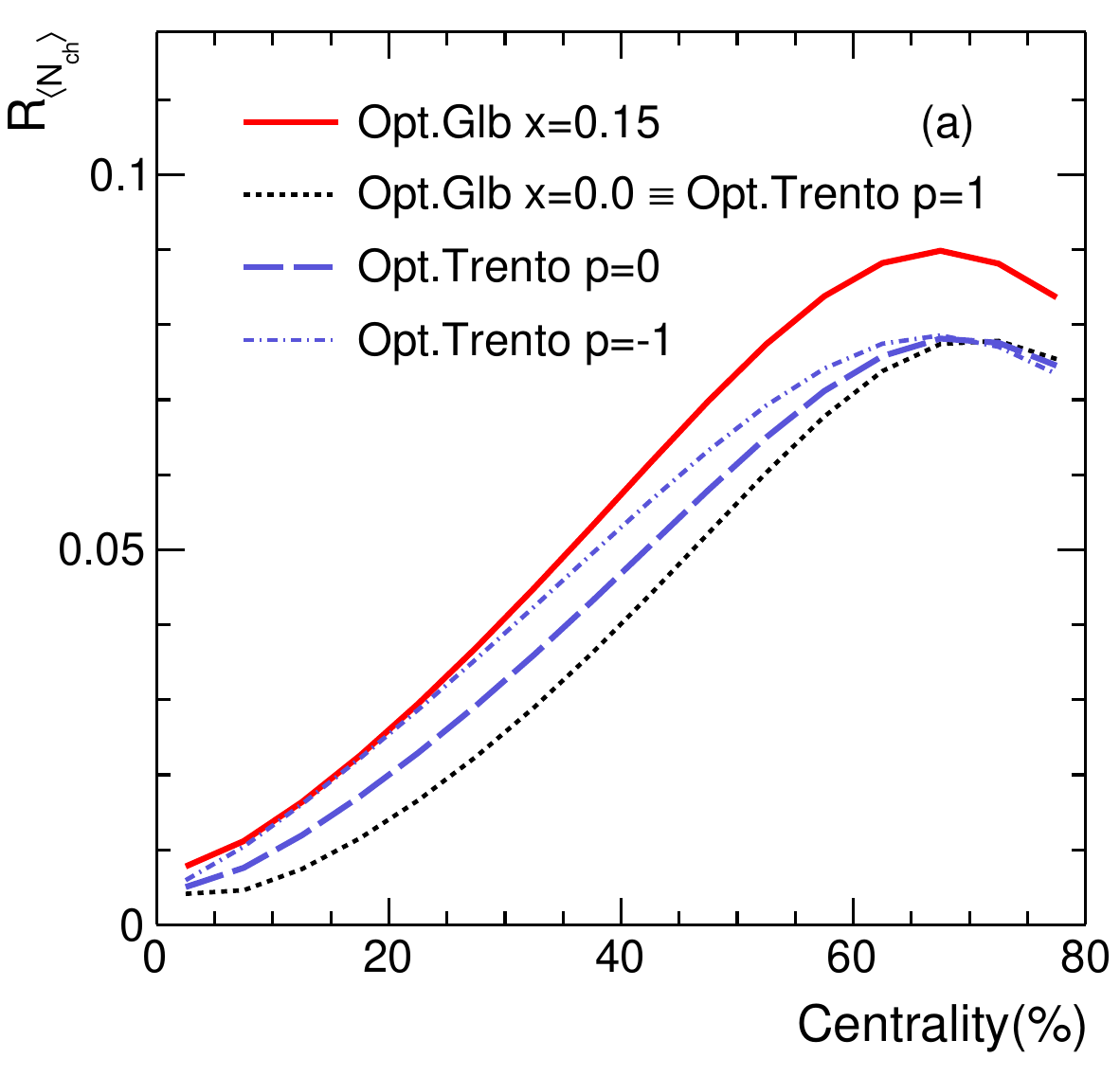}\includegraphics[scale=0.4]{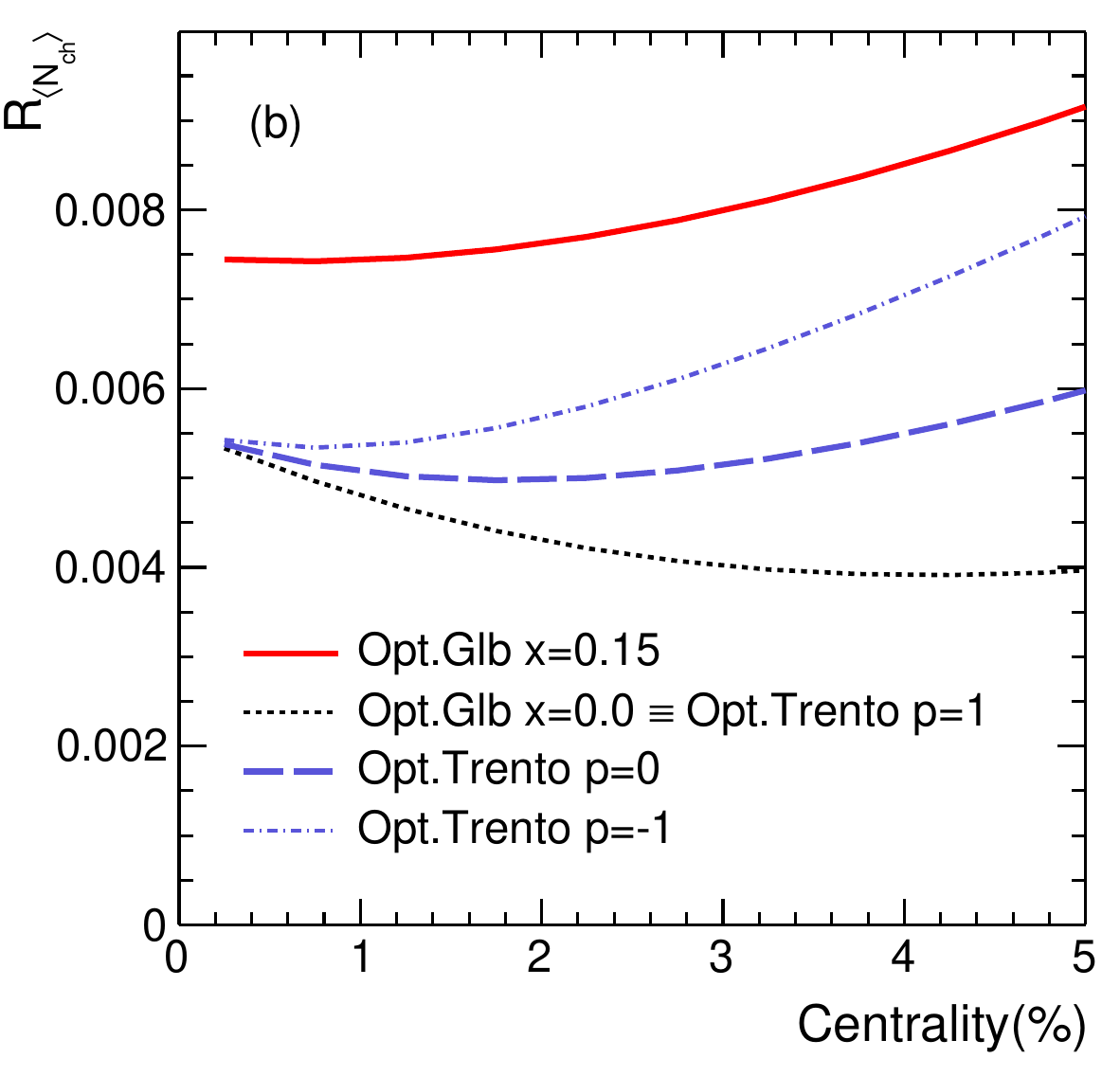}
      \caption{(Color online).
	 The centrality dependence of $\RNch$ from Optical Glauber (Opt.Glb) simulations with $x=0.15$ and $x=0$, and from Optical Trento (Opt.Trento) simulations with $p=0$ and $p=-1$ for (a) $0-80\%$ centrality range and (b) top $5\%$ centrality.
       \label{fig:optical1}} 
 \end{figure*}

The $\RNch$ obtained from the Optical Glauber model with $x=0$ and $x=0.15$, as well as those from the Optical Trento model with $p=0$ and $p=-1$, are shown in  Fig~\ref{fig:optical1}. The case of the Optical Glauber simulation with $x=0$ is nothing but the Optical Trento simulation with $p=1$, and the multiplicity, in this case, is only proportional to the number of participants. The $\RNch$ distribution depends on the model parameters $x$ and $p$. The magnitudes increase with $x$ and $p$ over the whole centrality range.  
For the most central collisions, the $\RNch$ increases with $x$ in Optical Glauber simulations, but is independent of $p$ in Optical Trento simulations. 
This is shown in the top $5\%$ centrality range in Fig.~\ref{fig:optical1}(b). 
The magnitudes of $\RNch$ converge to the same value from Optical Trento simulations with different values of $p$,
but the trends in centrality differ.

The parameters $x$ and $p$ are usually constrained by the individual multiplicity distributions (in this case, of isobar collisions or their mean).
The $\RNch$ in the most central collisions has been proposed to probe the neutron skin and symmetry energy. However, the particle production mechanisms will introduce uncertainties as shown in previous study~\cite{Li:2019kkh}.
The results shown in Fig.~\ref{fig:optical1} imply that the model uncertainty of $\RNch$ is negligible if the particle production follows the Trento model. 
They also suggest that the precision measurements of the mean multiplicity distribution of \RuRu\ and \ZrZr\ collisions and their difference may offer insights on the particle production mechanisms in relativistic heavy ion collisions.

 \begin{figure}[hbt!]
      \includegraphics[scale=0.4]{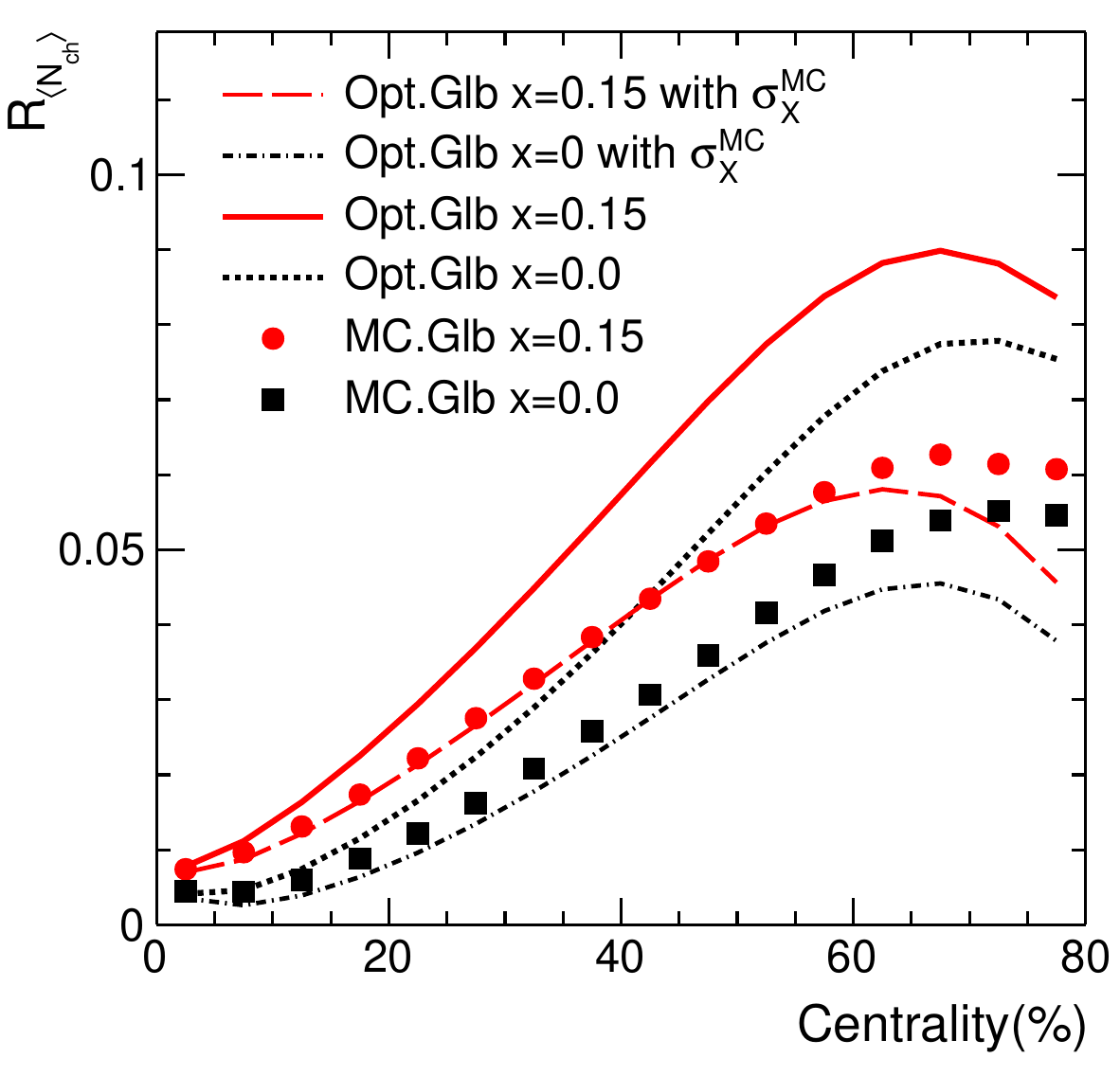}
      \caption{(Color online).
    The centrality dependence of $\RNch$ from Optical Glauber (Opt.Glb, solid curves) and Monte Carlo Glauber (MC.Glb, symbols) simulations with $x=0.15$ and $x=0$. The dashed curves are the results from Optical Glauber calculations with the total cross section from Mont Carlo Glauber simulations.
      \label{fig:optical2}} 
 \end{figure}

To study effects from event-by-event fluctuations, we present the $\RNch$ distributions from Monte Carlo Glauber simulations with $x=0.15$ and $x=0$ in Fig.~\ref{fig:optical2}. 
The apparent large fluctuation effect (i.e. difference in $\RNch$ between the Optical and the Monte Carlo Glauber) is because of  the different
total cross sections predicted by these two models~\cite{Miller:2007ri}.
For instance, the impact parameter cut would be smaller for a given centrality when a smaller total cross-section is used. 
Based on the nuclear density profiles in our study, the total cross sections for \RuRu\  and \ZrZr\ collisions are $\sigma_{X}^{\rm MC,RuRu} = 4.206$ barn and $\sigma_{X}^{\rm MC,ZrZr} = 4.348$ barn, which are $4.46\%$ and $3.73\%$ smaller than those in the Optical cases, respectively. 
To remove this trivial difference, we calculate the optical model using the total cross sections from the Monte Carlo Glauber. The results are shown in Fig.~\ref{fig:optical2};
they become similar now. Our results indicate that both the data of centrality dependent $\RNch$ and the model prediction of total cross section are important for the model to data comparison in relativistic isobar collisions~\cite{STAR:2021mii}.

\begin{figure*}[hbt!]
       \includegraphics[scale=0.4]{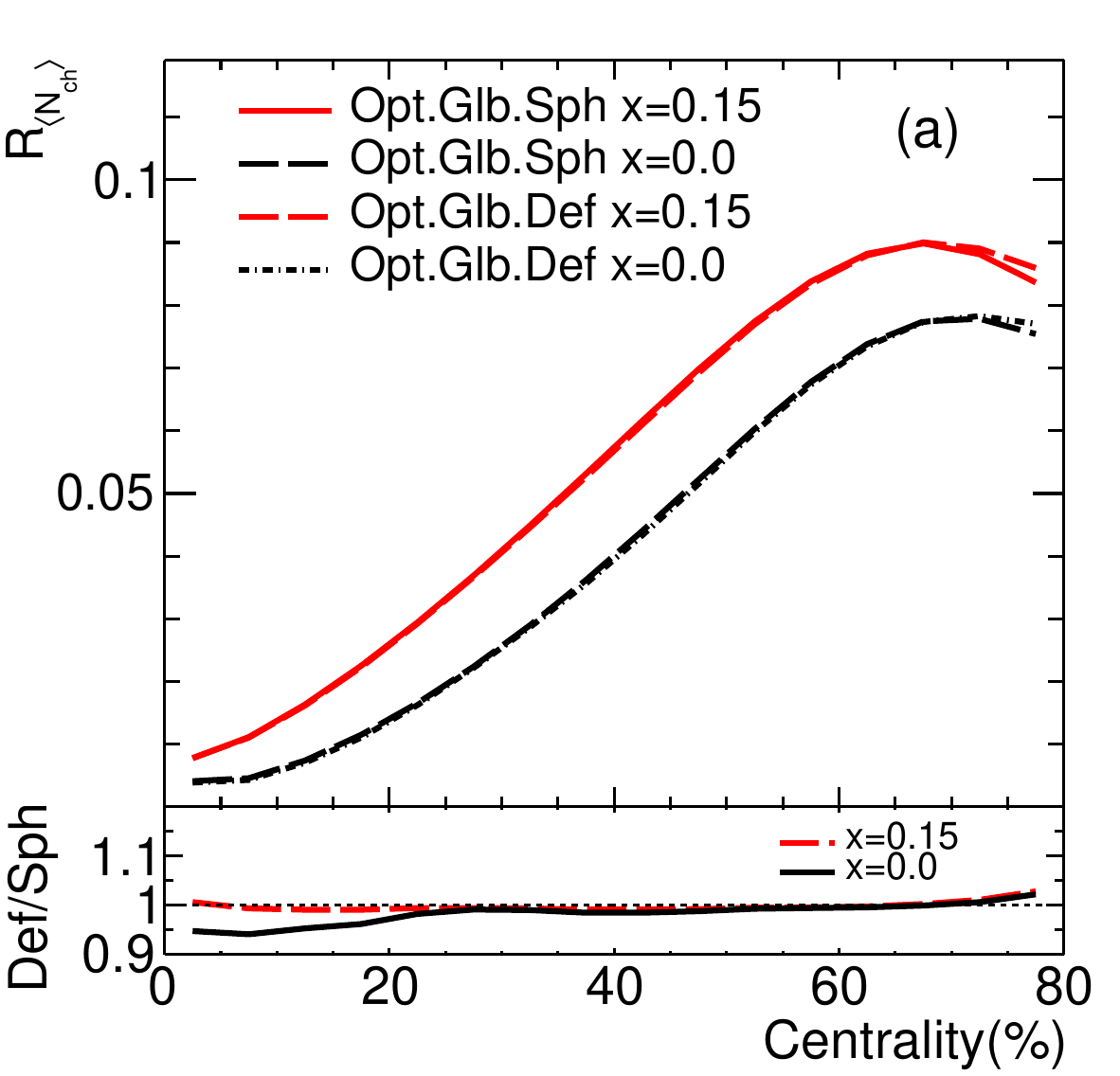}\includegraphics[scale=0.4]{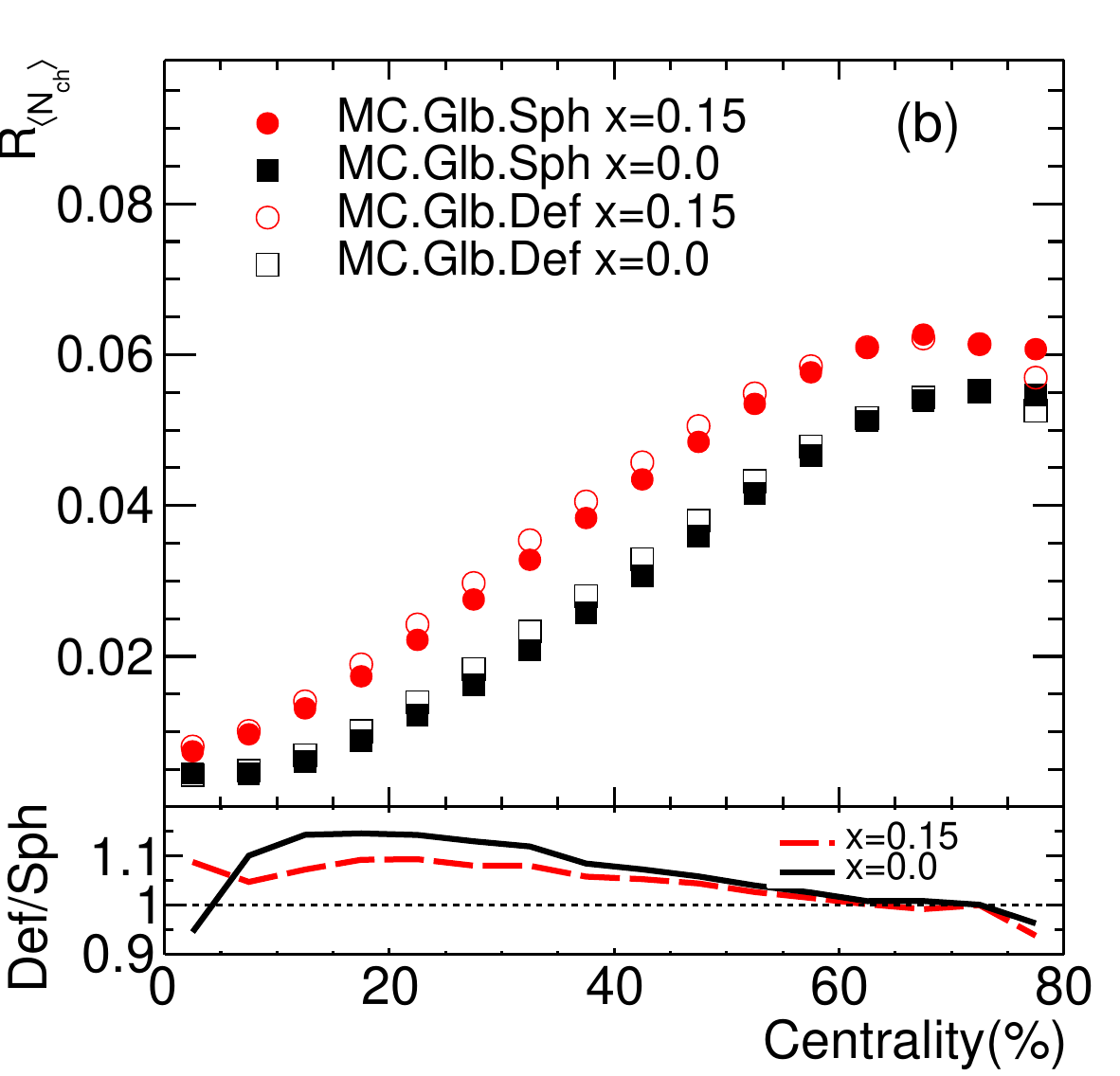}
      \caption{(Color online).
        The centrality dependence of $\RNch$ with spherical (Sph) and deformed (Def) colliding nuclei, calculated with (a) Optical Glauber (Opt.Glb) (b) Monte Carlo (MC.Glb) Glauber models.
      \label{fig:optical3}} 
 \end{figure*}

The $^{96}_{44}$Ru and $^{96}_{40}$Zr nuclei are deformed~\cite{Pritychenko:2013gwa,KIBEDI:2002wxc}, as also suggested by recent studies~\cite{STAR:2021mii,Zhang:2021kxj}.
The quadrupole deformation parameter for \Ru\ is $\beta_{2,Ru} = 0.16$ and is negligible for \Zr. 
The octupole deformation parameter is negligible for \Ru\ and is $\beta_{3,Zr} = 0.20$ for \Zr.  
Figure~\ref{fig:optical3} shows the $\RNch$ from deformed isobar collisions, 
where the volume and RMS radius of the colliding nuclei are constrained to match those of spherical nuclei~\cite{Xu:2021qjw,Zhao:2022uhl}. 
For both the Optical and Monte Carlo calculations, the results from deformed isobar collisions are similar to  those from spherical ones. Here we have used the same impact parameters listed in Tab~\ref{tab:Opt} for deformed isobar collisions. As  mentioned above, the centrality definition in the deformed case may be biased by multiplicity variations at a given impact parameter. However, the centrality  is well defined in the Monte Carlo Glauber simulation, and the consistency in  the corresponding results between spherical and deformed cases indicates that our centrality treatment in the Optical Glauber simulation is appropriate. The $\RNch$ is considerably larger in the Optical case because of the aforementioned larger total cross section. 
The differences are $\sim10\%$ between with and without nuclear deformation.
Based on previous study~\cite{Li:2019kkh}, this would result in an uncertainty on the extracted slope parameter of symmetry energy of about $5$ MeV. 
Our results indicate that, with appropriate magnitudes of deformation parameters, the effect of nuclear deformations on the uncertainties of $\RNch$ distributions are small. 

\begin{figure*}[hbt!]
      \includegraphics[scale=0.4]{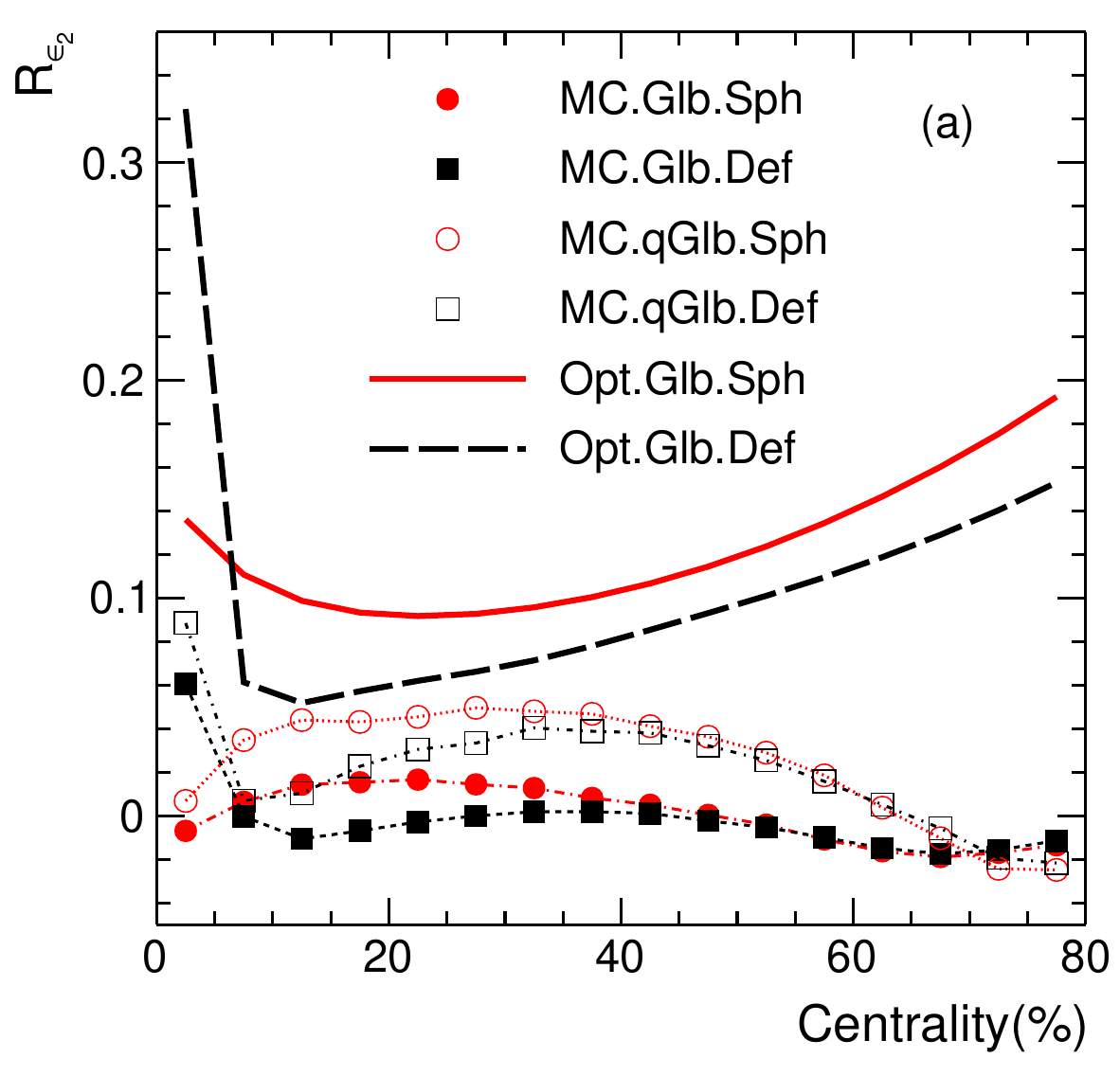} \includegraphics[scale=0.4]{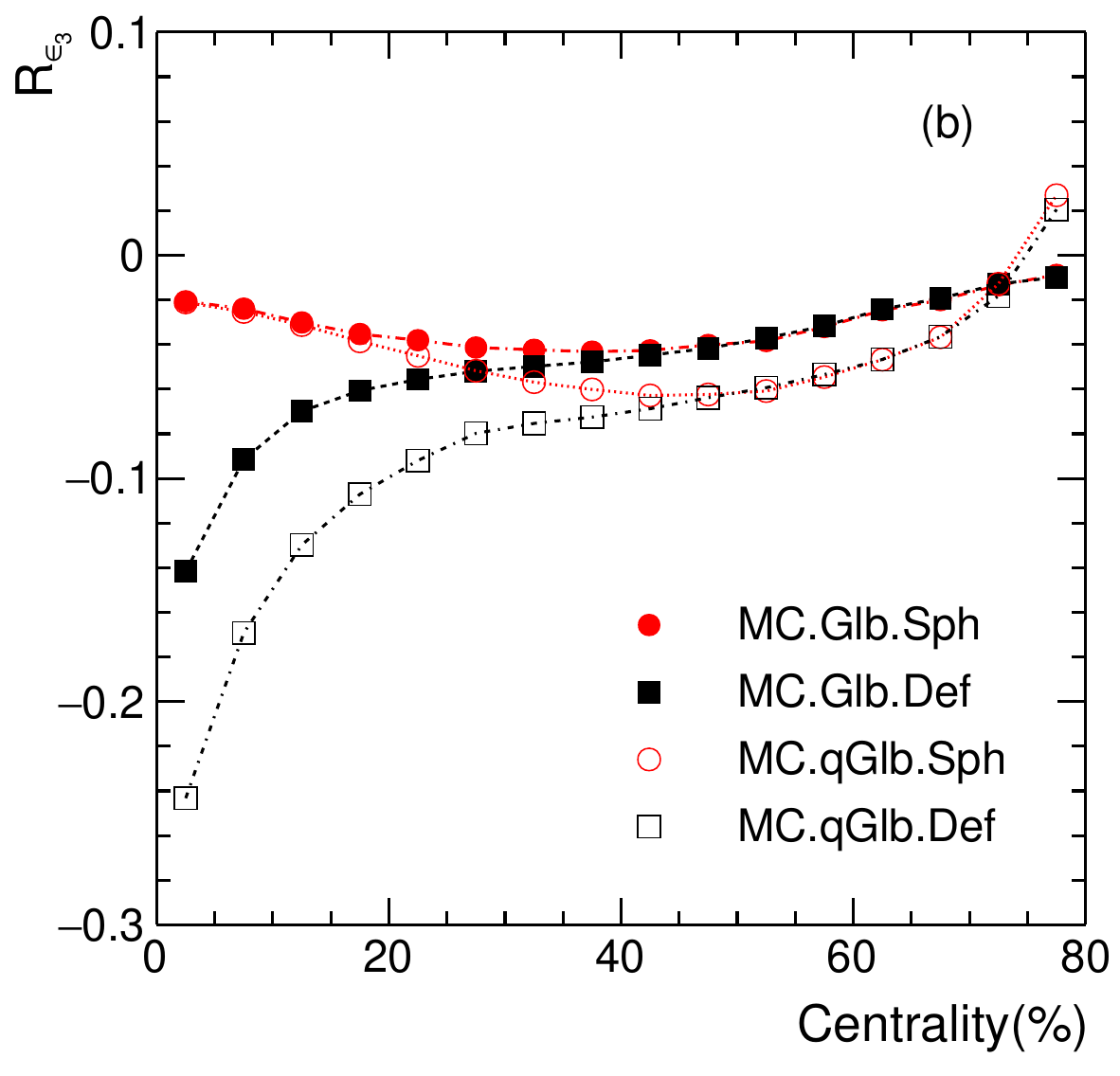}
      \caption{(Color online).
        The centrality dependence of (a) $\Ren$ and (b) $\Reo$. The results are calculated with spherical (Sph) and deformed (Def) colliding nuclei using Optical Glauber (Opt.Glb) and Monte Carlo Glauber (MC.Glb) models. 
        The results from Monte Carlo Glauber model with quark participant assumption (MC.qGlb) are presented as open symbols. From Optical Glauber simulations, $\epsilon_{3}$ vanishes due to symmetry in spherical cases, and the $\Reo<-1$ in deformed cases are outside the frame of the plot.
      \label{fig:optical4}} 
 \end{figure*}

This is not the case for the anisotropic flow observables, where previous studies suggest that both nuclear deformation and event-by-event fluctuations are important~\cite{Miller:2007ri,Alver:2010gr,Schenke:2010rr}. Apart from  nuclear deformation, most of the anisotropic flow produced in the  most central heavy-ion collisions is due to the position fluctuations of  nucleons in the colliding nuclei. 
In this paper, we study the effect of nuclear deformations and initial fluctuations on geometry eccentricities $\Reno$ (n=2,3). The results are shown in Fig.~\ref{fig:optical4}. 
Note that, for the Optical Glauber simulations, $\epsilon_{3}$ vanishes   for the spherical case due to symmetry, and for the deformed case the $\Reo<-1$ (because $\epsilon_{3}^{\rm ZrZr}\gg\epsilon_{3}^{\rm RuRu}$)  is outside the frame of Fig.~\ref{fig:optical4}(b).  
Comparing the results between  deformed  and spherical nuclei, the general features are similar for both. This is consistent with the results shown in previous studies~\cite{Zhang:2021kxj,Jia:2022qgl}, where the large quadrupole deformation $\beta_{2}$ in Ru increases the $\Ren$ in most central collisions while the large octupole deformation $\beta_{3}$ suppresses the values in mid-central collisions. 

The magnitudes of $\Ren$ show significant differences between the Optical and Monte Carlo Glauber models. The $\Ren$ caused by the nuclear geometry is significantly diluted by the event-by-event fluctuations. The reason is rather straightforward: event-by-event fluctuations contribute to a large fraction of the  eccentricities in Monte Carlo simulations, and these contributions are common between \RuRu\ and \ZrZr\ collisions, suppressing the differences caused by nuclear deformation and/or collision geometry. 

In the Monte Carlo Glauber simulation, the participating nucleons are usually treated as point-like particles as we did in the simulations above. However,  nucleons have substructures~\cite{Eremin:2003qn,Loizides:2016djv,Albacete:2017ajt}. More substructure constituents  are expected to suppress the fluctuations. We use the quark participant assumption as it was done in Ref.~\cite{Eremin:2003qn},  increasing the density by three times and decreasing the inelastic cross-section by a factor of $9$. The results are also shown in Fig.~\ref{fig:optical4}. The Monte Carlo Glauber simulations under quark participant scenario significantly increase the $\Ren$ and decrease $\Reo$, bringing the values closer to the Optical limit. We note that despite of the cross-section differences, the $\Reno$ calculated in the Optical Glauber model are similar to those calculated in the Monte Carlo Glauber model with respect to reaction plane instead of participant plane~\cite{Xu:2017zcn}, which is more related to the elliptic flow measurement with respect to zero-degree calorimeter (ZDC) instead of time projection chamber (TPC) in experiment~\cite{STAR:2021mii}.

There are more sources of initial fluctuations in heavy ion collisions, for example, the correlations between the quarks in a nucleon, as well as the correlations between the nucleons in the colliding nuclei. We find that the minimum distance between two nucleons in simulations, as one of the sources of nucleon-nucleon correlations, does not change our results. 
We note that the final-state effects, which are not the focus of the present study, are also important for the accurate extraction of the symmetry energy slope parameter. We postpone such investigation to a future work.

\section{Summary}\label{sec:summary}
In this work, the effects of nuclear deformations and initial fluctuations on the
ratio observables in relativistic isobar collisions
are studied using the Optical Glauber and Monte Carlo Glauber models.
The GPU parallel computing technology is used in our calculations to achieve high precision, especially for the deformed colliding nucleus. 
Our main findings are as follow.
\begin{itemize}
	\item $\RNch$ depends on particle production mechanisms and model parameters, while it converges to a single value in most central collisions in the Trento model (see Fig. 1). 
	\item Initial fluctuations affect the total cross section and thus the $\RNch$, however the effect is negligible for most central collisions (see Fig.~2).
	\item Because of the constraints of the total volumes and the measured RMS radii of the isobar nuclei, their deformations are found to have negligible effects on $\RNch$ (see Fig.~3).
	\item The $\Ren$ and $\Reo$, driven by the nuclear diffusenesses and deformations, are suppressed by the initial finite number fluctuations, dependent on the degree of freedom of nucleons or constituent quarks (see Fig.~4). 
\end{itemize}
Our results indicate unique (in)sensitivities of the ratio observables of $\RNch$ and $\Reno$ to nuclear deformations and initial fluctuations in relativistic isobar collisions.
These features, compared to data, may potentially probe the particle production mechanism and the physics underlying nuclear structure.
\section*{Acknowledgments}
This work is supported in part by the National Natural Science Foundation of China under Grant Nos.~12275082, 12035006, 12075085 (HX), the Zhejiang Provincial Natural Science Foundation of China under Grant No.~LY21A050001 (HX), 
and the U.S.~Department of Energy under Grant No.~DE-SC0012910 (FW). The authors are grateful to the C3S2 computing center of Huzhou University for its calculation support.  
\bibliography{ref2}
\end{document}